\documentclass[fp,twocolumn]{jpsj3}
\usepackage{txfonts}
\usepackage{amsmath,amssymb,bm,comment,here}

\title{Study on Susceptibilities of Superconductors BaPtSb and BaPtAs with Honeycomb Structure}

\author{Naoya \textsc{Furutani}$^1$, Yoshiki \textsc{Imai}$^2$\thanks{imai@ous.ac.jp}, Tsuyoshi \textsc{Imazu}$^3$, and Jun \textsc{Goryo}$^3$}
\inst{$^1$Department of Applied Physics, Okayama University of Science, 1-1 Ridaicho, Kita-ku, Okayama-shi, Okayama 700-0005, Japan \\
$^2$Department of Physics, Okayama University of Science, 1-1 Ridaicho, Kita-ku, Okayama-shi, Okayama 700-0005, Japan \\
$^3$Department of Mathematics and Physics, Hirosaki University, Aomori 036-8561, Japan} 

\abst{
The low-energy electronic properties of the new superconductors BaPtSb and BaPtAs with an ordered honeycomb network are investigated in the normal phase, where the former compound is a candidate for time-reversal symmetry-breaking superconductors. 
By means of the first-principles calculation, we show that there exist two-dimensional cylinder-like Fermi surfaces around the $k_z$ axis and one outer spherelike Fermi surface around the K and K$^{'}$ points in both compounds, which are mainly composed of Pt 5$d$ and Sb 5$p$/As 4$p$ electrons.
We construct low-energy effective models, which are well described by using three bands consisting of two Pt 5$d$ orbitals and one Sb $5p$/As $4p$ orbital. 
By evaluating susceptibilities using effective models, we find that dominant contributions to those susceptibilities result from the outer spherelike Fermi surface.  
Whereas out-of-plane fluctuations are enhanced in both compounds in the higher-temperature region, in-plane fluctuations become dominant in the very low-temperature region in BaPtSb owing to a better nesting condition in the $k_z=0$ plane from the outer spherelike Fermi surface. 
These fluctuations yield instabilities to ordered states, such as superconductivity, and might be associated with the occurrence of the superconducting state with time-reversal symmetry breaking in BaPtSb.
}

\begin{document}
\maketitle

\section{Introduction}

Time-reversal symmetry-breaking (TRSB) superconductors have attracted much interest as prime candidates for topological superconductivity~\cite{qi2011,bansil2016,sato2017}. 
In particular, the lattice structure with the honeycomb-like network has an advantage in the occurrence of TRSB superconductivity ~\cite{uchoa2007,black-schaffer2007,honerkamp2008,kuroki2010,kiesel2012,wan2012,black-schaffer2012}. The hexagonal KZnAs-type compound SrPtAs shows superconductivity with $T_c\sim 2.4$ K~\cite{nishikubo2011} with the locally noncentrosymmetric lattice and the space group $P6_3/mmc$ (No. 194). A muon spin relaxation measurement ($\mu$SR) study showed the appearance of an intrinsic spontaneous magnetic field in the superconducting phase~\cite{biswas2013}, which indicated the TRSB superconducting state. Whereas studies with nuclear quadrupole resonance measurement (NQR)~\cite{matano2014} and magnetic penetration depth measurement~\cite{landaeta2016} suggested the conventional $s$-wave pairing, theoretical investigations showed that there would be room for considering the possibility of chiral $d$-wave pairing with TRSB.~\cite{fischer2014,fischer2015,ueki2019}

Recently, superconductivities of the new hexagonal compounds BaPtSb ($T_c \sim 1.64$ K) and BaPtAs ($T_c \sim 2.8$ K) have been reported. The lattice structure of both compounds is of the hexagonal SrPtSb type with the space group $P\bar{6}m2$ (No. 187) without an inversion center.~\cite{kudo2018a,kudo2018b}.  In a study on the solid solution BaPt (As$_{1-x}$Sb$_x$), the transition temperature shows non-monotonic behavior, whereas the lattice parameters increase linearly and the symmetry is unchanged with respect to the doping ratio $x$~\cite{ogawa2022}. 
Moreover, a preliminary $\mu$SR study on BaPtSb suggested the appearance of a spontaneous magnetic field~\cite{adachi2018}. 
These results imply that the pairing symmetry is changed by varying $x$; and thus, it would be important to clarify the difference between the microscopic properties of BaPtSb and BaPtAs.

In the present study, we investigate the low-energy electronic properties of BaPtSb and BaPtAs with the space group $P\bar{6}m2$ by using the first-principles method in the normal phase, and construct effective model Hamiltonians with the tight-binding approach. By evaluating static susceptibilities using effective models, we identify the difference between the electronic structures of BaPtSb and BaPtAs, and discuss the interplay between dominant fluctuations and the low-energy electronic properties in detail. 

This paper is organized as follows. The electronic structures obtained by the first-principles method are given in the next section. In Sec. 3, we construct effective model Hamiltonians on the basis of the low-energy electronic structures obtained in Sec. 2. The static susceptibilities are discussed in Sec. 4. A summary is given in Sec. 5.

\section{Electronic Structures from the First-principles Method}
In this section, we discuss the electronic structures of BaPtSb and BaPtAs with the space group $P\bar{6}m2$, for which calculations are carried out
by using the Quantum Espresso Package based on the plane wave pseudopotential method~\cite{giannozzi2009,giannozzi2017}. The k-point mesh is employed up to $24 \times 24 \times 16$ ($48 \times 48 \times 32$) in the Brillouin zone (BZ) in scf (nscf) calculations, and the energy cutoff of wavefunctions is set to 100 Ryd. The lattice constants are given by $a = 4.535$ \AA\, and $c = 4.884$ \AA\, for BaPtSb~\cite{wenski1986}, and $a = 4.308$ \AA\, and $c = 4.761$ \AA\, for BaPtAs\cite{kudo2018b}. 

\begin{figure}[tb]
	\centering
	\includegraphics[width=42mm]{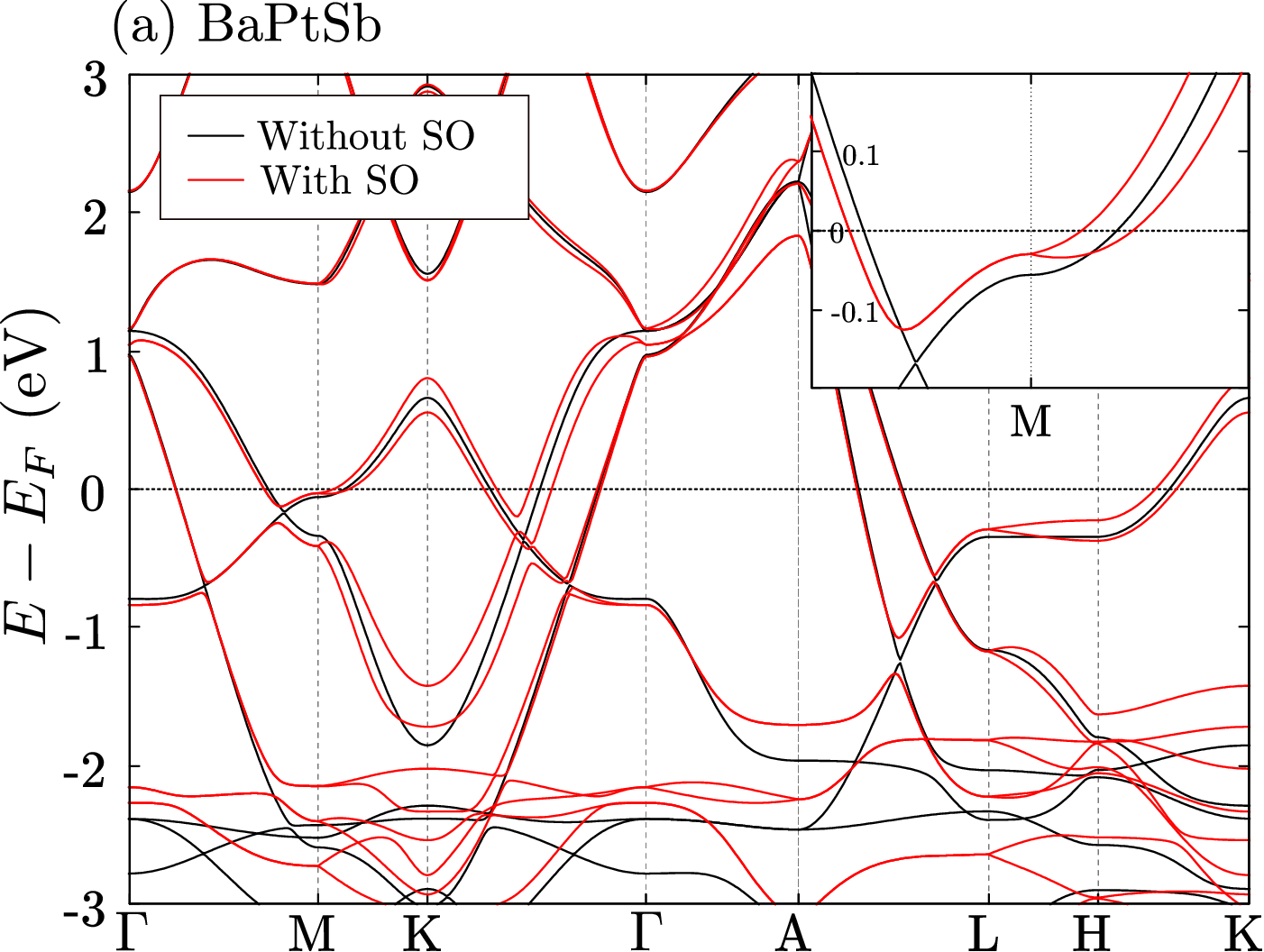}
	\includegraphics[width=42mm]{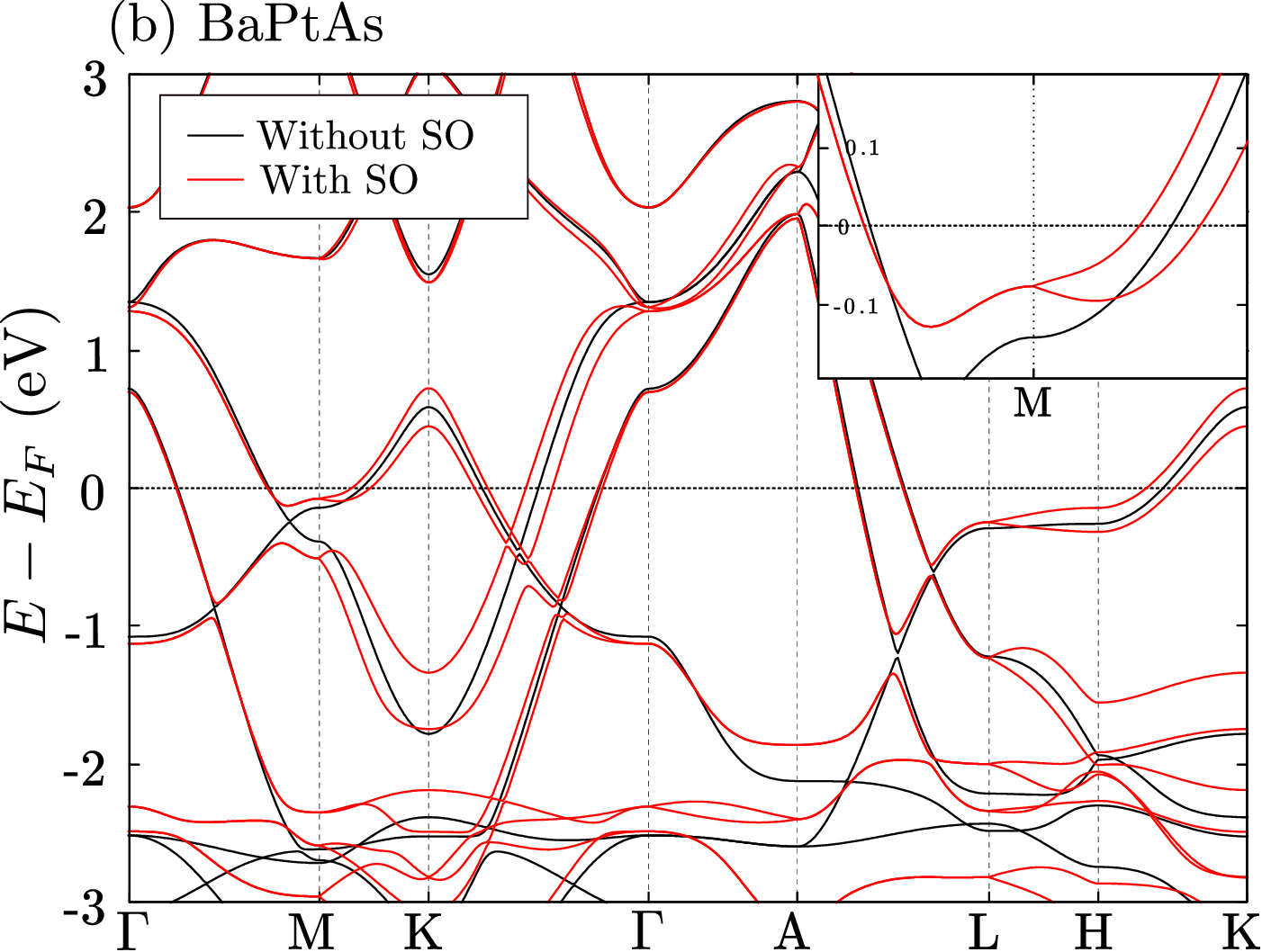}
	\caption{(Color online) Electronic structures obtained by the first-principles calculation; (a) BaPtSb and (b) BaPtAs. The Fermi levels are set to zero. The black (red) lines denote results without (with) the SO interaction. The insets show the magnifications in the low-energy region around the M point. }
	\label{bands}
\end{figure}

The obtained electronic structures are shown in Fig. \ref{bands}. 
While the overall shapes of structures in both compounds are generally identical, the energy eigenvalues of the highest valence states around the M point in BaPtSb are very close to the Fermi level in comparison with those in BaPtAs. 
Since the M point corresponds to the van Hove singularity (vHS), the slope of the energy dispersion around the M point near the Fermi level becomes almost flat.
Thus, electron velocities are suppressed strongly, and the density of states (DOS) increases.  
The spin-orbit (SO) interaction induces the splitting of energy eigenvalues on M$-$K$-$$\Gamma$ and L$-$H lines, and is notable in the in-plane direction.  

\begin{figure}[b]
	\centering
	\includegraphics[width=75mm]{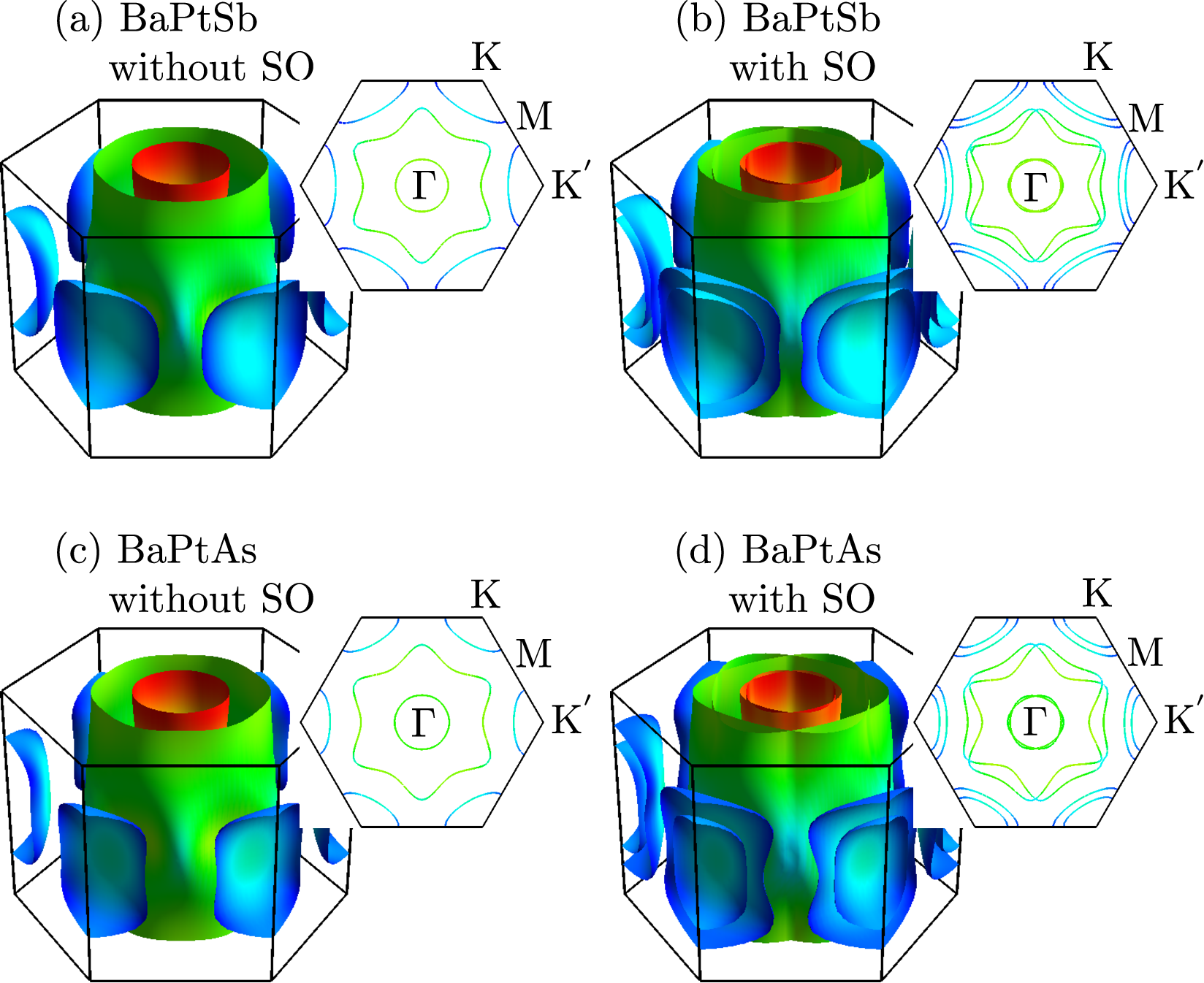}
	\caption{(Color online) Fermi surfaces; (a,b) BaPtSb and (c,d) BaPtAs without/with the SO interaction (drawn using fermisurfer~\cite{kawamura2019}). The warm (cool) color denotes the large (small) amplitude of electron velocities.  The insets show sections of the Fermi surfaces in the $k_z=0$ plane. }
	\label{FS}
\end{figure}

Figure \ref{FS} shows Fermi surfaces of BaPtSb and BaPtAs without/with the SO interaction. 
There exist three (six) Fermi surfaces without (with) the SO interaction with the two-dimensional cylinder-like Fermi surfaces around the $k_z$ axis and the three-dimensional spherelike Fermi surface around the K and K$^{'}$ points. 
Since the low-energy electronic structures of BaPtSb around the M point are very close to the Fermi level, the presence of external fields and/or perturbations can lead to a change in topology around the M point in the outer (spherelike) Fermi surface. 

\begin{figure}[tb]
	\centering
	\includegraphics[width=70mm]{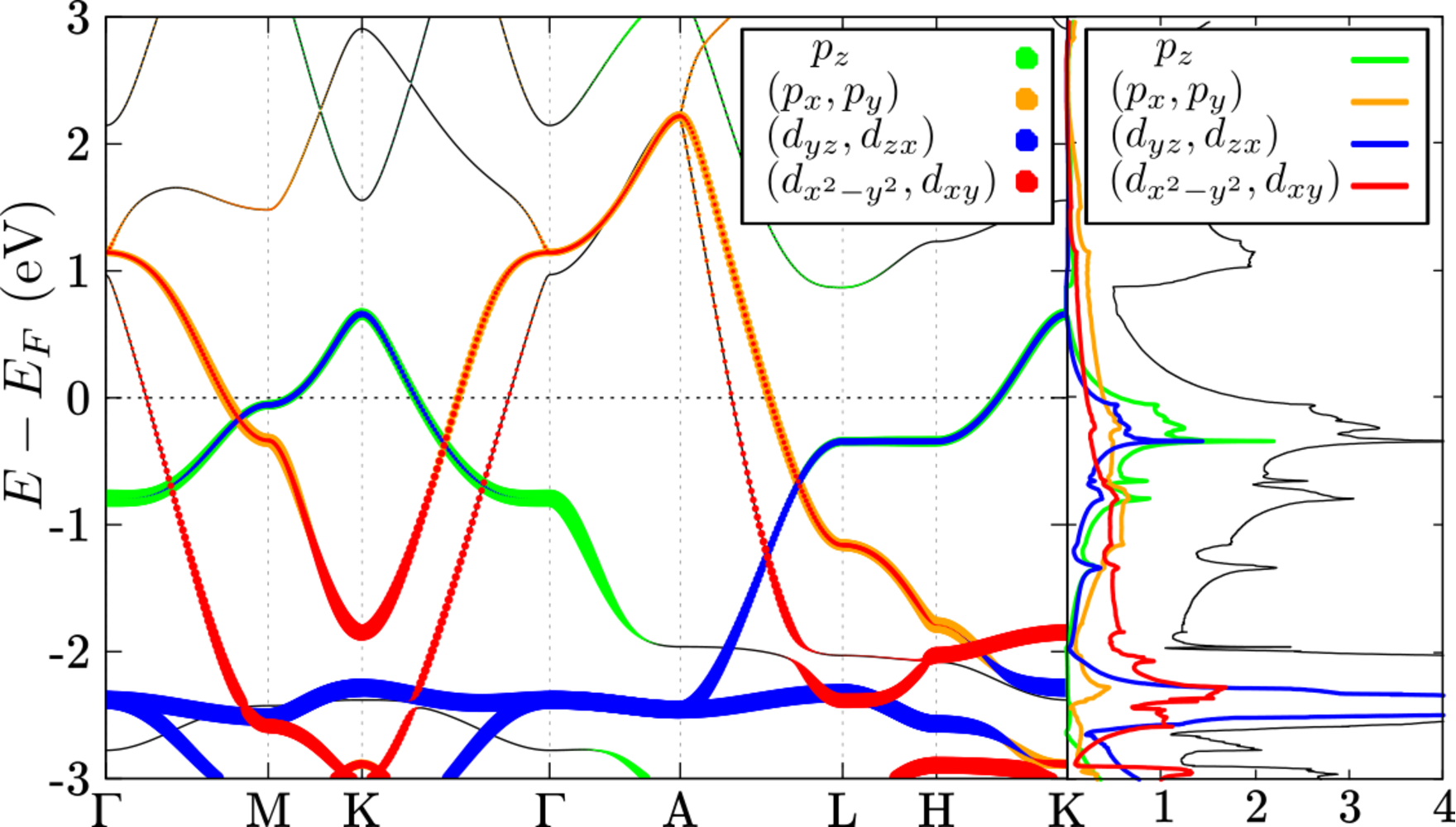}
	\caption{(Color online) The left (right) panel represents weights of orbitals in the electronic structure (total and partial DOSs) of BaPtSb without the SO interaction. }
	\label{fatbandplot}
\end{figure}

To identify the contribution of each orbital to the low-energy electronic structures, the projected weights of orbitals and the DOS of BaPtSb without the SO interaction are depicted in Fig. \ref{fatbandplot}. 
The electronic structures around the Fermi level are mainly governed by the Pt 5$d$ $x^2-y^2$, $xy$, $yz$, and $zx$ orbitals, and the Sb 5$p$ $x$, $y$, and $z$ orbitals, where the weights of the Pt 5$d$ and Sb 5$p$ orbitals are comparable.
The main part of energy eigenvalues of the Pt 5$d$ $3z^2-r^2$ orbital is far from the Fermi level, which is negligible. 

The projected weights of the orbitals show that the energy eigenvalues are divided into two groups, \{Pt 5$d$ ($x^2-y^2, xy$), Sb 5$p$ ($x,y$)\} and \{Pt 5$d$ ($yz,zx$), Sb 5$p$ ($z$)\}, where the energy eigenstates are composed of orbitals in the same groups.  
Note that a similar result is obtained for BaPtAs. 

\section{Construction of Effective Tight-binding Models}

Tight-binding models are a useful tool for microscopically investigating physical properties including the superconducting state. 
In this section, on the basis of the results obtained by the first-principles method discussed in the previous section, we construct effective tight-binding models to describe the low-energy electronic structures of BaPtSb and BaPtAs by using the Slater-Koster (SK) method~\cite{slater1954}, where the models and the physical quantities obtained by this method exactly maintain the lattice symmetry.  

In the previous section, we showed that the low-energy electronic structures consist of two groups, \{Pt 5$d$ ($x^2-y^2, xy$), Sb 5$p$/As 4$p$ ($x,y$)\} and \{Pt 5$d$ ($yz,zx$), Sb 5$p$/As 4$p$ ($z$)\}. 
Although a strong hybridization exists between orbitals in the same groups, we extract Pt 5$d$ $x^2-y^2, xy$, and Sb 5$p$/As 4$p$ $z$ orbitals as bases. 

Let us construct effective three-band tight-binding models by using the SK method. 
Lattice structures are defined by the following primitive translation vectors $\bm a_i$  $(i=1\sim 3)$ and positions $\bm r_\nu$ ($\nu=$Pt, Sb/As) in the unit cells:
\begin{align}
	&\bm a_1=\left(a,0,0 \right),\>\bm a_2=\left(-\frac{a}{2},\frac{\sqrt{3}a}{2},0 \right),\>\bm a_3=\left(0,0,c \right),\\
	&\bm r_{\text{Pt}}=\left(\frac{a}{2},\frac{a}{2\sqrt{3}},0 \right), \>\bm r_{\text{Sb/As}}=\left(0,0,0 \right), 
\end{align}
where $a$ ($c$) stands for the lattice constant in the $a$-$b$ plane (along the $c$ axis). 
$a$ and $c$ are taken to be unity for simplicity. 

Then, the Hamiltonian is written as 
\begin{align}
	H&=\sum_{\bm k}\sum_{m,m'}\sum_{\sigma,\sigma'}\varepsilon^{(m\sigma,m'\sigma')}_{\bm k}c^\dag_{\bm k m\sigma}c_{\bm km'\sigma'},\label{eq:hml1}\\
	\varepsilon^{(m\sigma,m'\sigma')}_{\bm k}&=\sum_{i,j}t_{ij}^{mm'}e^{-i\bm k\cdot(\bm R_i-\bm R_j+\bm r_m-\bm r_{m'})}\delta_{\sigma\sigma'}+ \epsilon_{m}\delta_{mm'}\delta_{\sigma\sigma'},  
	\label{eq:hml2}
\end{align}
where $c^\dag_{im\sigma}$ $(c_{im\sigma})$ is the creation (annihilation) operator for electrons on site $i$ and the orbital $m$ (Pt 5$d$ $\pm 2$ and Sb 5$p$ /As 4$p$ $z$) with the spin $\sigma$. The spherical harmonics are employed as bases with respect to Pt 5$d$ orbitals ($|m\rangle =|\pm 2\rangle=|x^2-y^2\rangle \pm i |xy \rangle $) around the quantization axis ($z$-axis), so that the hopping amplitudes $t_{ij}^{mm'}$ are independent of the hopping direction. 
$\bm R_i$, $\bm k$, and $\epsilon_{m}$ denote the position of the unit cell, the wave vector, and the on-site potential, respectively. 

The electronic structures obtained by the first-principles calculation show that the SO interaction affects the dispersion relations, especially in the in-plane direction. Since the electronic structures of BaPtSb and BaPtAs are classified by the irreducible representation with the point group $D_{3h}$, we introduce the following form as the SO term: 
\begin{align}
	\lambda^{(m\sigma,m'\sigma')}_{\bm k}=i\sigma \lambda_m \sum_{l=1}^3\sin (\bm k \cdot \bm T_l)\delta_{mm'}\delta_{\sigma\sigma'}, 
	\label{eq:so}
\end{align}
where the orbitally diagonalized form is assumed for simplicity. $\bm T_l$ stands for the nearest-neighbor vector in the $a$-$b$ plane between the same atoms with $\{\bm T_1, \bm T_2, \bm T_3\} = \{\bm a_1, \bm a_2, -(\bm a_1+\bm a_2) \}$. $\lambda_m$ denotes the amplitude of the SO interaction in the orbital $m$, and $\sigma$ corresponds to $1(-1)$ for the spin $\uparrow(\downarrow)$. This equation is added to Eq. (\ref{eq:hml2}).

The hopping amplitudes $t_{ij}^{mm'}$ are given by the SK parameters. The hopping range is considered between the nearest-neighbor (NN), next- nearest-neighbor (NNN), and 3rd-nearest-neighbor (3NN) sites. 
Here, we emphasize that we also examine the Wannier90 method~\cite{pizzi2020}, and by referring to the results of this method, we determine the amplitudes to reproduce particularly the low-energy structures, as shown in Table \ref{tbl:sk}. 
\begin{table}[tb]
	\begin{flushleft}
	\begin{tabular}{c||c|c}\hline
		
		 & onsite potential ($\lambda_m$=0) & onsite potential ($\lambda_m=0.02$) \\\hline
		Pt & $-1.3$ & $-1.3$ \\\hline
		Sb/As  & $-0.60$ & $-0.55/-0.57$ \\\hline
	\end{tabular}
	\end{flushleft}
	\begin{tabular}{l|ccc}
		\hline
		Pt--Pt & ($dd\sigma$)& ($dd\pi$)& ($dd\delta$)  \\\hline
		$a$-$b$ plane (NN)& $0.22/0.25$ & $0.72$ & $-0.40$ \\\hline
		$a$-$b$ plane (NNN) & $0.10$ & $-0.02$ & $-0.2$  \\\hline
		$a$-$b$ plane (3NN) & $-$ & $0.13$ & $-$ \\\hline
		$c$ axis (NNN) & $-$ & $0.10$ & $-0.10$ \\\hline
		$c$ axis (3NN) & $-$ & $-0.05$ & $-$ \\\hline \hline
		Sb--Sb/As-As& $(pp\sigma)$& $(pp\pi)$& \\\hline
		$a$-$b$ plane (NN)  & $-$  & $-0.17$ \\\hline
		$a$-$b$ plane (NNN) & $-$ & $0.02$ \\\hline
		$a$-$b$ plane (3NN) & $-$ & $0.00/0.01$  \\\hline\hline
		$c$ axis (NN)  & $0.185/0.12$ & $-$ \\\hline
		$c$ axis (NNN) & $0.04$ & $-$ \\\hline
		$c$ axis (3NN) & $0.06$ & $-$ \\\hline\hline
		Pt--Sb/As& $(pd\sigma)$& $(pd\pi)$ & \\\hline
		$c$ axis (NN)  &$ 0.70/1.10$  & $0.10/0.20$ \\\hline
		$c$ axis (NNN) &$-0.15$ & $-$ \\\hline
	\end{tabular}
	\caption{Onsite potentials without/with the SO interaction, amplitudes of the SO interaction, and hopping parameters in the $a$-$b$ plane and along the $c$-axis between Pt--Pt, Sb--Sb/As--As, and Pt--Sb/As sites. $-$ represents zero, and other components with zero amplitude up to 3NN sites are not shown. The numbers before and after $"A/B"$ correspond to BaPtSb and BaPtAs, respectively. The energy unit is in eV. }
	\label{tbl:sk}
\end{table}
$(\alpha \beta m)$ represents the hopping amplitude depending only on the distance between the sites with the $\alpha m$ and $\beta m$ orbitals. 
Except for onsite potentials, we assume that parameters are independent of the SO interaction for simplicity. 
The hopping between the Pt and Sb/As sites in the $a$-$b$ plane vanishes owing to the lattice symmetry.

\begin{figure}[tb]
	\centering
	\includegraphics[width=70mm]{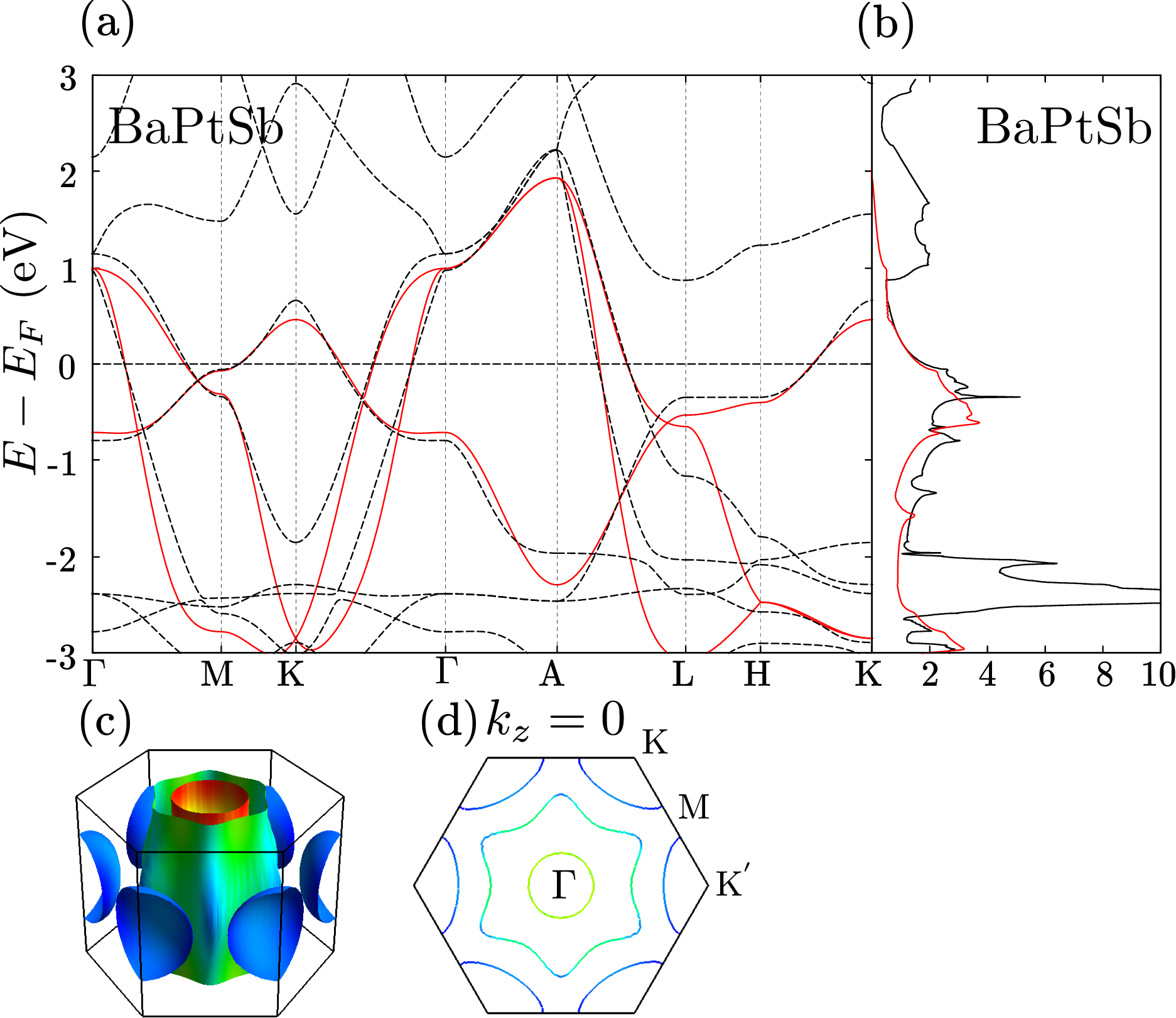}
	\includegraphics[width=70mm]{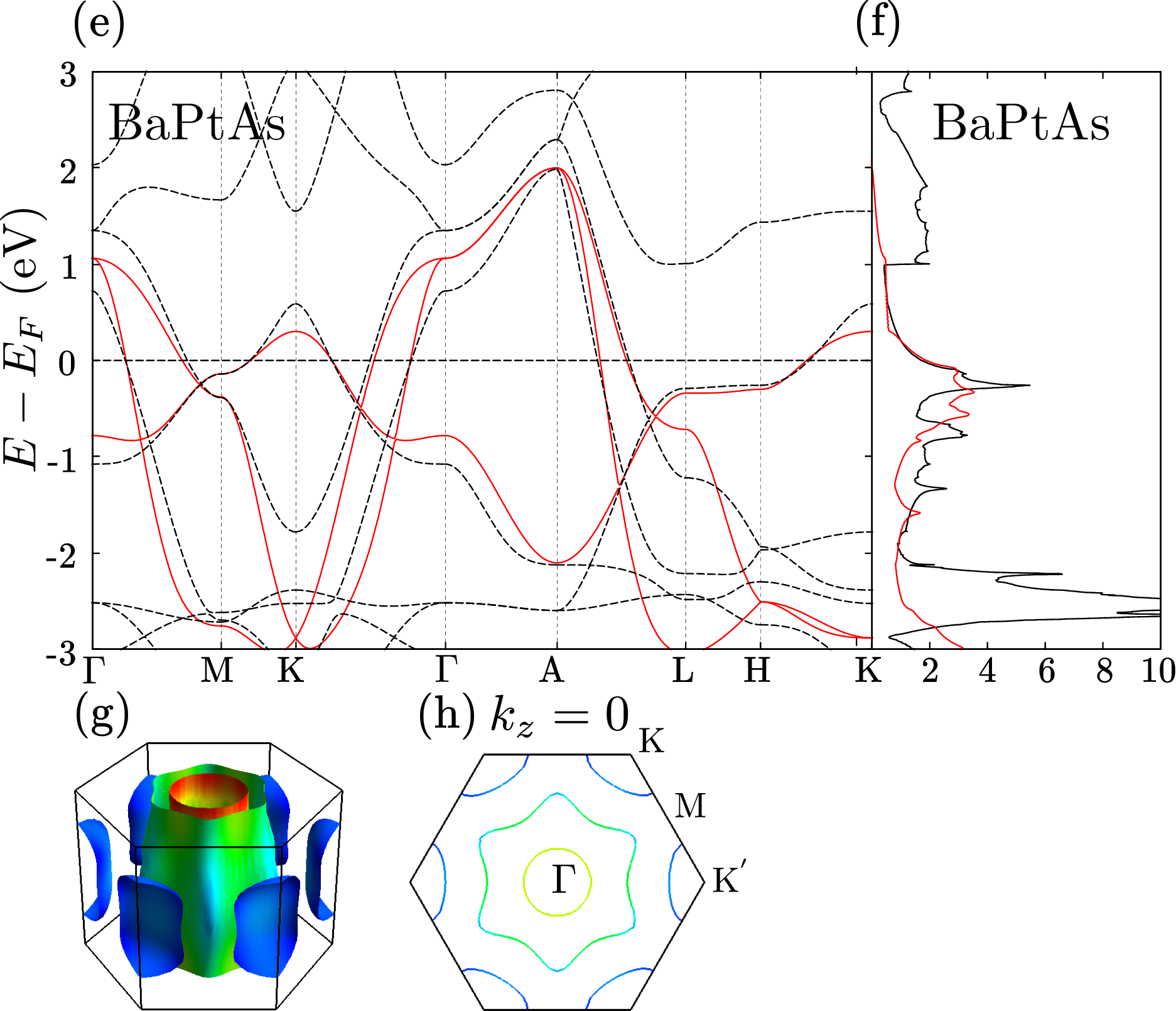}
	\caption{(Color online) Electronic structures, DOSs, Fermi surfaces, and sections of Fermi surfaces in the $k_z=0$ plane without the SO interaction obtained by the SK method; (a--d) BaPtSb and (e--f) BaPtAs. (a,b,e,f) The solid red (dashed black) lines denote results obtained by the SK method (first-principles calculation). The Fermi levels are set to zero. 
	}
	\label{SKbands}
\end{figure}

By diagonalizing the Hamiltonian in Eq. (\ref{eq:hml1}), we obtain the energy eigenvalues of BaPtSb and BaPtAs, which are depicted in Fig. \ref{SKbands}.  
The electronic structures reproduce the first-principles ones qualitatively. In particular, the Fermi surface structures coincide in the $k_z=0$ plane.  
This result indicates that the low-energy electronic structures are described well by  three-band models and that the effects of the eliminated orbitals are renormalized partially in the hopping parameters. 

\begin{figure}[b]
	\centering
	\includegraphics[width=85mm]{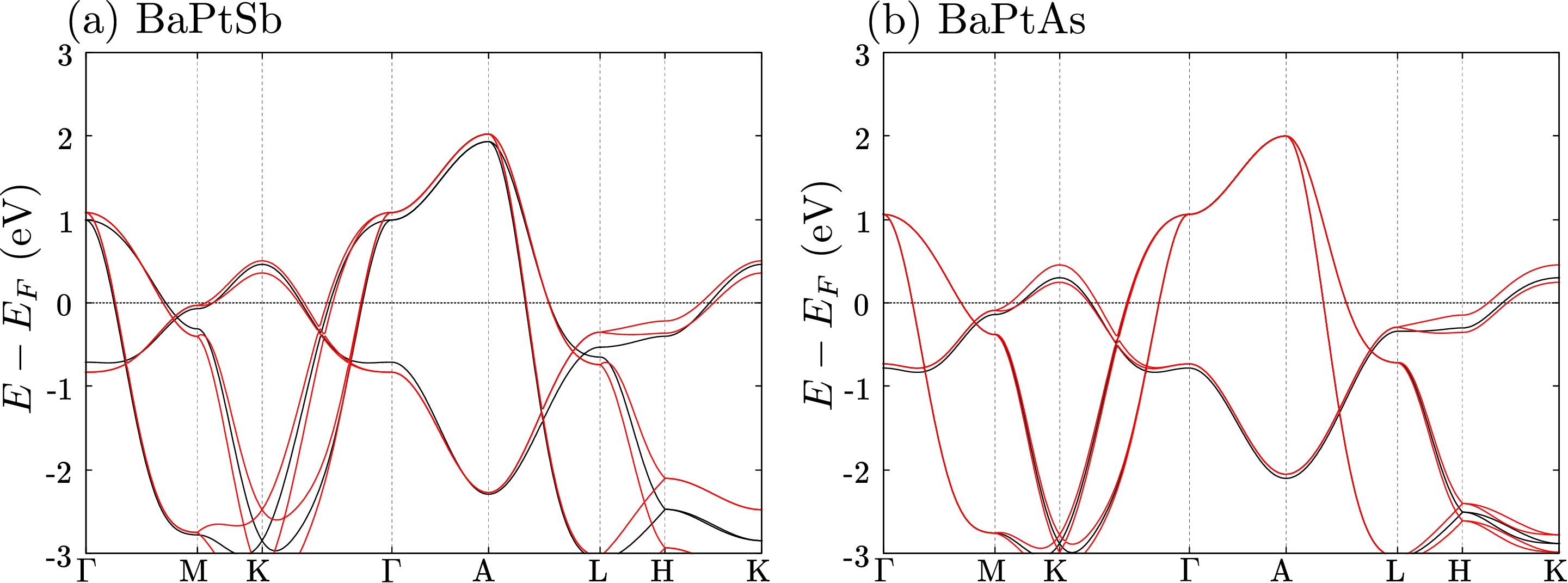}
	\caption{(Color online) Electronic structures without/with the SO interaction obtained by the SK method; (a) BaPtSb and (b) BaPtAs. The Fermi levels are set to zero. The black (red) lines denote results without (with) the SO interaction. }
	\label{SKbandsSO}
\end{figure}

The low-energy electronic structures including the SO interaction are depicted in Fig. \ref{SKbandsSO}. 
The overall shapes of electronic structures and wave vectors corresponding to the splitting of electronic structures are in agreement with the first-principles results.  

\section{Static Susceptibilities}
By calculating static susceptibilities in the low-temperature region using the effective models constructed in the previous section, we discuss the difference between BaPtSb and BaPtAs concerning dominant fluctuations and the instabilities to ordered states.  

\subsection{Definition of Susceptibilities}
To investigate fluctuation effects, the irreducible susceptibility $\hat{\chi}$ with the matrix form is defined, whose matrix component at the wave vector $\bm q$ is given by 
\begin{align}
	\chi_{m_1m_2,m_3m_4}^{\sigma_1\sigma_2,\sigma_3\sigma_4}(\bm q, i\omega_n)&=\int_0^\beta d\tau \,e^{i\omega_n \tau}\langle \rho_{m_1m_2}^{\sigma_1\sigma_2}(\bm q,\tau)\rho^{\dag\sigma_3\sigma_4}_{m_3m_4}(\bm q,0) \rangle, \\
	\rho_{mm'}^{\sigma\sigma'}(\bm q,\tau)&=\frac{1}{\sqrt{N}}\sum_ie^{i\bm q \cdot \bm R_i}c_{im\sigma}^\dag(\tau)c_{im'\sigma'}(\tau). 
\end{align}
$\rho_{mm'}^{\sigma\sigma'}(\bm q,\tau)$ is the generalized polarization at the imaginary time $\tau$~\cite{imai2003}. 
$\omega_n$ represents the bosonic Matsubara frequency and $\beta$ is the inverse temperature. 

Assuming a weak coupling regime, we employ random the phase approximation (RPA) approach, where a matrix component in the irreducible susceptibility $\hat{\chi}$ is written with the non-interacting Green's function $G_{mm'}^{\sigma\sigma'}(\bm k, i\omega_l)$ as
\begin{align}
	\chi_{m_1m_2,m_3m_4}^{\sigma_1\sigma_2,\sigma_3\sigma_4}(\bm q, i\omega_n)&=-\frac{1}{\beta N}\sum_{\bm k,l}G_{m_2m_4}^{\sigma_2\sigma_4}(\bm k+\bm q,i\omega_n+i\omega_l)\nonumber\\
	&\hspace{-15mm}\times G_{m_3m_1}^{\sigma_3\sigma_1}(\bm k,i\omega_l)e^{(\bm k+\bm q)\cdot (\bm r_{m_2}-\bm r_{m_4})}e^{-\bm k\cdot (\bm r_{m_1}-\bm r_{m_3})}, \\
	G_{mm'}^{\sigma\sigma'}(\bm k, i\omega_l)&=\left.\left[i\omega_l\hat{I}-\hat{\varepsilon}_{\bm k} \right]^{-1}\right|_{mm'}\delta_{\sigma\sigma'},   
\end{align}
where $\hat{I}$ and $\omega_l$ stand for the unit matrix and fermionic Matsubara frequency, respectively. 
$\hat{\varepsilon}_{\bm k}$ is a matrix with the summation of the kinetic energy in Eq. (\ref{eq:hml2}) and the SO term in Eq. (\ref{eq:so}). 
The irreducible susceptibility is in the $9\times 9$ ($36\times 36$) matrix form without (with) the SO interaction. 
Note that to evaluate the accuracy of the effective models constructed by the SK method, we have also calculated the irreducible susceptibilities using the hopping amplitudes obtained by the Wannier90 method.

The response functions are described with the irreducible susceptibility within RPA. The spin and charge susceptibilities are defined by
	$
	\hat{\chi}^{(s,c)}(\bm q, i\omega_n)=\hat{\chi}(\bm q, i\omega_n)\left[\hat{I}\mp \hat{U}\hat{\chi}(\bm q, i\omega_n) \right]^{-1}, 
	$
where the $-(+)$ sign denotes the spin (charge) susceptibility. 
The maximum values of these susceptibilities in the first BZ exhibit dominant fluctuations, which are associated with the instabilities to ordered states.
 
In the weak coupling regime, the spin and charge susceptibilities directly reflect the feature of irreducible susceptibility. 
We have confirmed that by assuming the electron--electron interaction to be a Hubbard-like one with an orbitally diagonal form written with
	$
	U_{m_1m_2,m_3m_4}^{\sigma_1\sigma_2,\sigma_3\sigma_4}=U\delta_{m_1,m_2}\delta_{m_3,m_4}\delta_{m_1,m_3}\delta_{\sigma_1,\sigma_3}\delta_{\sigma_2,\sigma_4}\delta_{\sigma_1,\bar{\sigma_2}},    
	$
the largest eigenvalues of the spin (charge) susceptibility are enhanced (suppressed) further than those of the irreducible susceptibility, while maintaining the overall shapes in the whole BZ.  
Thus, to discuss dominant fluctuations, we focus on the irreducible susceptibility hereafter.  

\subsection{Largest Eigenvalues of Static Susceptibilities}
\begin{figure*}[tb]
	\centering
	\includegraphics[width=160mm]{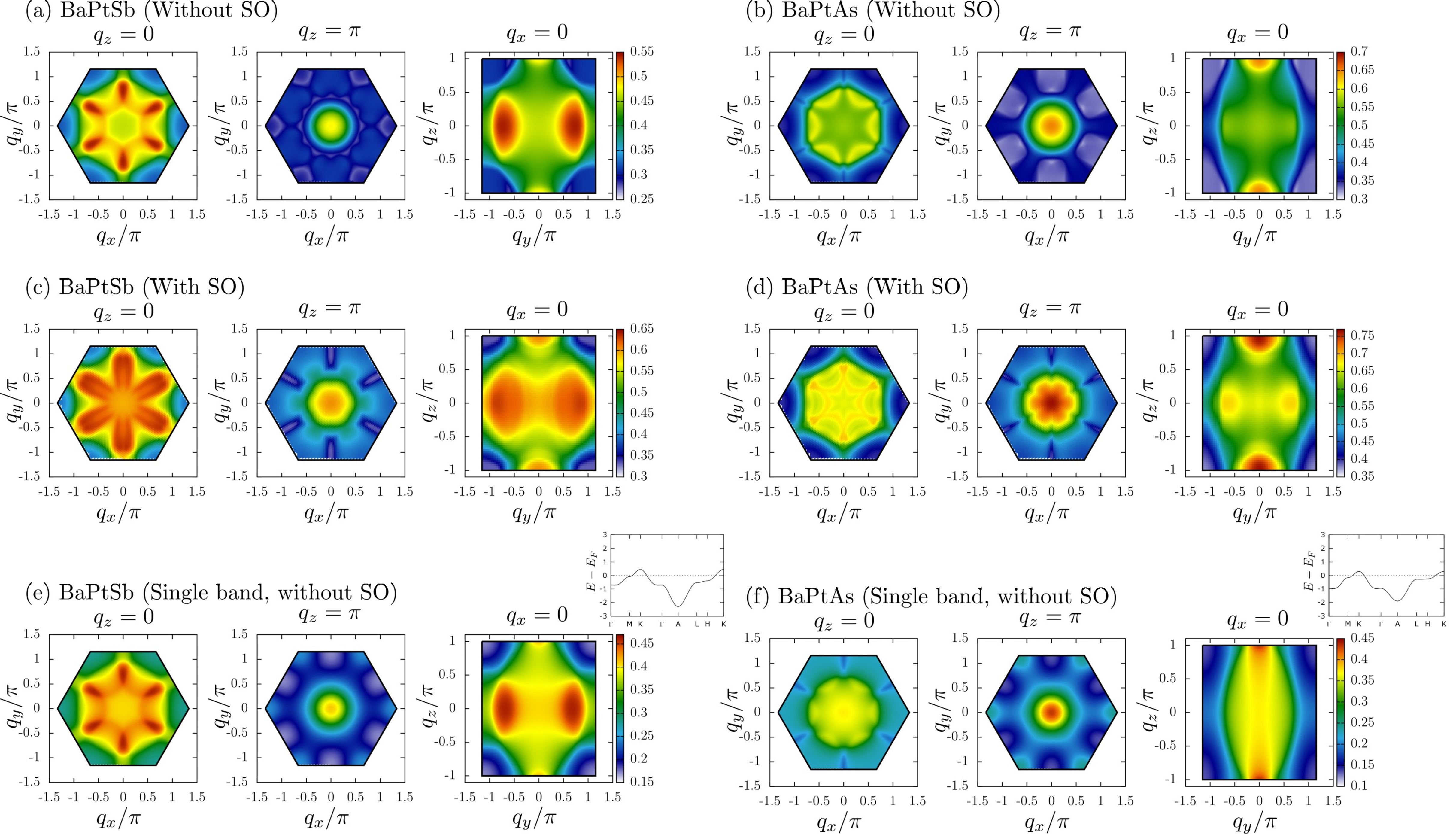}
	\caption{(Color online) Largest eigenvalues of static susceptibilities at sections with respect to several $\bm q$ points in the first BZ. The left (a,c,e) and right (b,d,f) panels represent the results of BaPtSb and BaPtAs, respectively. 
	The top (a,b), middle (c,d), and bottom (e,f) panels correspond to the results without the SO interaction, with the SO interaction, and of the single-band model without the SO interaction, respectively. 
	The temperature is $k_BT=0.01$ eV. 
	The insets in panels (e,f) represent energy dispersions of single-band models. 
	}
	\label{chi}
\end{figure*}

Figures \ref{chi} (a) and \ref{chi} (b) show the largest eigenvalues of static irreducible susceptibilities without the SO interaction at low-temperature with $k_BT=0.01$ eV, which are calculated from the effective three-band models constructed by the SK method. 
Whereas the peaks are at $\bm q\simeq (0,0,\pi)$ in both compounds, the maximum peaks in BaPtSb at $k_BT=0.01$ eV appear on the $q_x=0$ line and its equivalent lines with the lattice symmetry in the $q_z=0$ plane. 
In contrast, the maximum value appears in the out-of-plane direction with $\bm q\simeq (0,0,\pi)$ in BaPtAs in the whole temperature range. 
These results indicate that dominant fluctuations in BaPtSb and BaPtAs differ in the very  low-temperature region. 

Figures \ref{chi} (c) and \ref{chi} (d) show the results with the SO interaction.  
Although the maximum values are altered slightly and the peak structures are smeared around the peak position $\bm q$, they are essentially similar to results without the SO interaction. 

The eigenstates corresponding to the largest eigenvalues of susceptibilities in the whole first BZ originate from the outer spherelike Fermi surfaces in both compounds. 
Thus, we also construct single-band models, describing the outer spherelike Fermi surfaces with the $p_z$ orbital, and evaluate static susceptibilities, whose largest eigenvalues are depicted in Figs. \ref{chi} (e) and (f).   
All results are  qualitatively in agreement with those from three-band models, which indicates that the prime contribution to dominant fluctuations results from the outer spherelike Fermi surfaces around the K and K$^{'}$ points.

\begin{figure}[b]
	\centering
	\includegraphics[width=70mm]{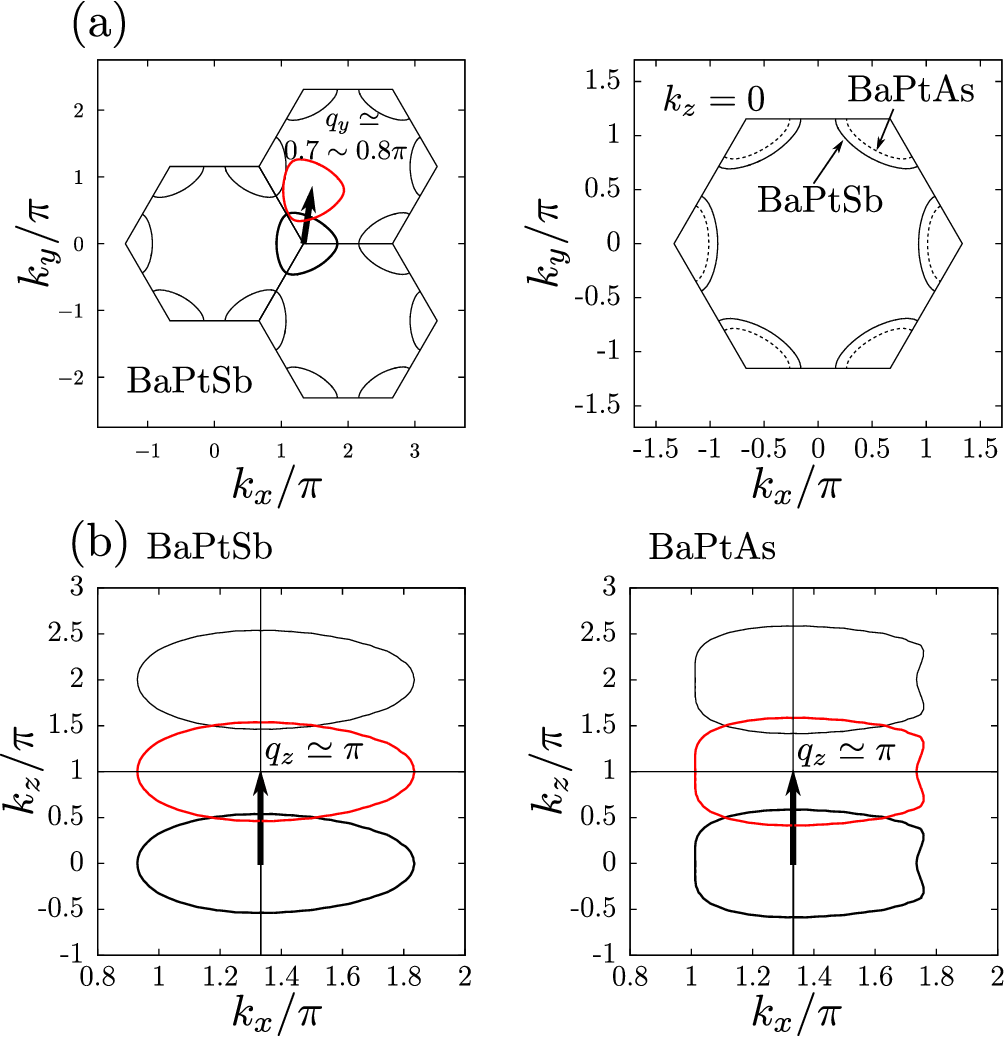}
	\caption{(Color online) Nesting of the outer spherelike Fermi surfaces.  Nesting vectors are defined by the shift of the Fermi surface (solid red line), which touches or overlaps the original at several $\bm q$ points with $D_{3h}$ symmetry. The examples of nesting vectors are indicated by bold arrows in the figures. (a)  (Left panel) Solid triangles represent the outer spherelike Fermi surfaces of BaPtSb in the $k_z=0$ plane in the extended BZ scheme. (Right panel) Difference of the outer spherelike Fermi surfaces between BaPtSb (solid line) and BaPtAs (dashed line). (b) Nesting vectors are depicted along the $k_z$ direction in the $k_y=0$ plane around the BZ boundary ($k_x=4\pi/3$) for  BaPtSb  and BaPtAs. }
	\label{nesting}
\end{figure}

The peak positions of static susceptibility are explained by the Fermi surface nesting. 
Figure \ref{nesting} shows the nesting of the outer Fermi surfaces of each compound around the K and K$^{'}$ points.
We see that, in BaPtSb, the cross section in the $k_z=0$ plane has a well-defined nesting vector, and the Fermi level is close to the vHS (shown in Fig.\ref{SKbands}). Thus, the static susceptibility of BaPtSb in the in-plane direction is enhanced around the nesting vectors in the low-temperature region. 

\begin{figure}[b]
	\centering
	\includegraphics[width=50mm]{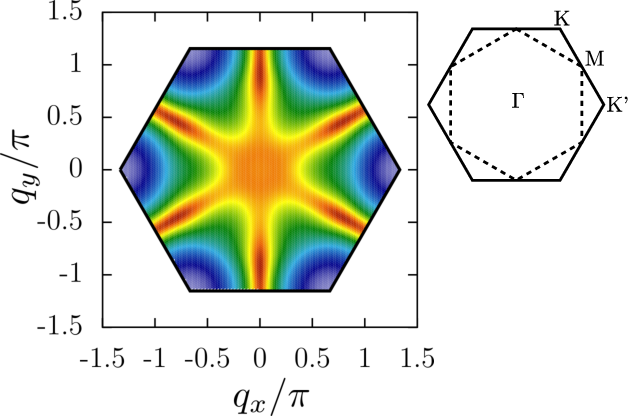}
	\caption{(Color online) Static susceptibility in the two-dimensional triangular lattice with $1/4$ doping at $k_BT=0.1t$, where $t$ is the nearest-neighbor hopping amplitude. The inset shows the Fermi surface (dashed line) in the BZ. }
	\label{trilat_chi}
\end{figure}

This result is similar to those obtained from the two-dimensional triangular and honeycomb lattice systems with $1/4$ and $1/8$ doping with the nearest-neighbor hopping. 
In both lattice systems with doping, the Fermi level is on the vHS and the Fermi surfaces become a regular hexagon with good nesting condition. 
Figure \ref{trilat_chi} displays the susceptibility in the two-dimensional triangular lattice with $1/4$ doping. 
The peak position of the susceptibility appears around the M point and changes slightly depending on temperature. This result is similar to susceptibilities in BaPtSb in the $k_z=0$ plane shown in Figs. \ref{chi} (a), (c), and (e). 
These systems have an advantage in the occurrence of a TRSB superconducting state such as the $d_{x^2-y^2}+id_{xy}$ state~\cite{uchoa2007,black-schaffer2007,honerkamp2008,kuroki2010,kiesel2012,wan2012,black-schaffer2012,honerkamp2003,xiaodong2018}. 
Thus, there is a potential for the occurrence of the TRSB superconducting state in BaPtSb. 

Since there is no strong nesting in the $k_z=0$ plane in BaPtAs whose Fermi level is away from the vHS in comparison with the BaPtSb case, the static susceptibility of BaPtAs is suppressed in the in-plane direction. 
As the Fermi level moves away from the vHS, the in-plane nesting condition becomes worse. 
Thus, this result might be associated with the study on the solid solution BaPt(As$_{1-x}$Sb$_x$), where the suppression of $T_c$ occurs with decreasing $x$ around $x=1$~\cite{ogawa2022}. 

The Fermi surface nesting also exists along the $k_z$ direction.  
In both compounds, each Fermi surface touches itself with the nesting vector $\bm q\simeq (0,0,\pi)$, whose contributions become dominant in the higher-temperature region. In contrast to the BaPtSb case, the dominant mode in BaPtAs becomes $\simeq (0,0,\pi)$ even in the low-energy region, where the pairing symmetry in the superconducting state might be with the horizontal node. 

\subsection{Temperature Dependence of Static Susceptibilities}
\begin{figure}[b]
	\centering
	\includegraphics[width=40mm]{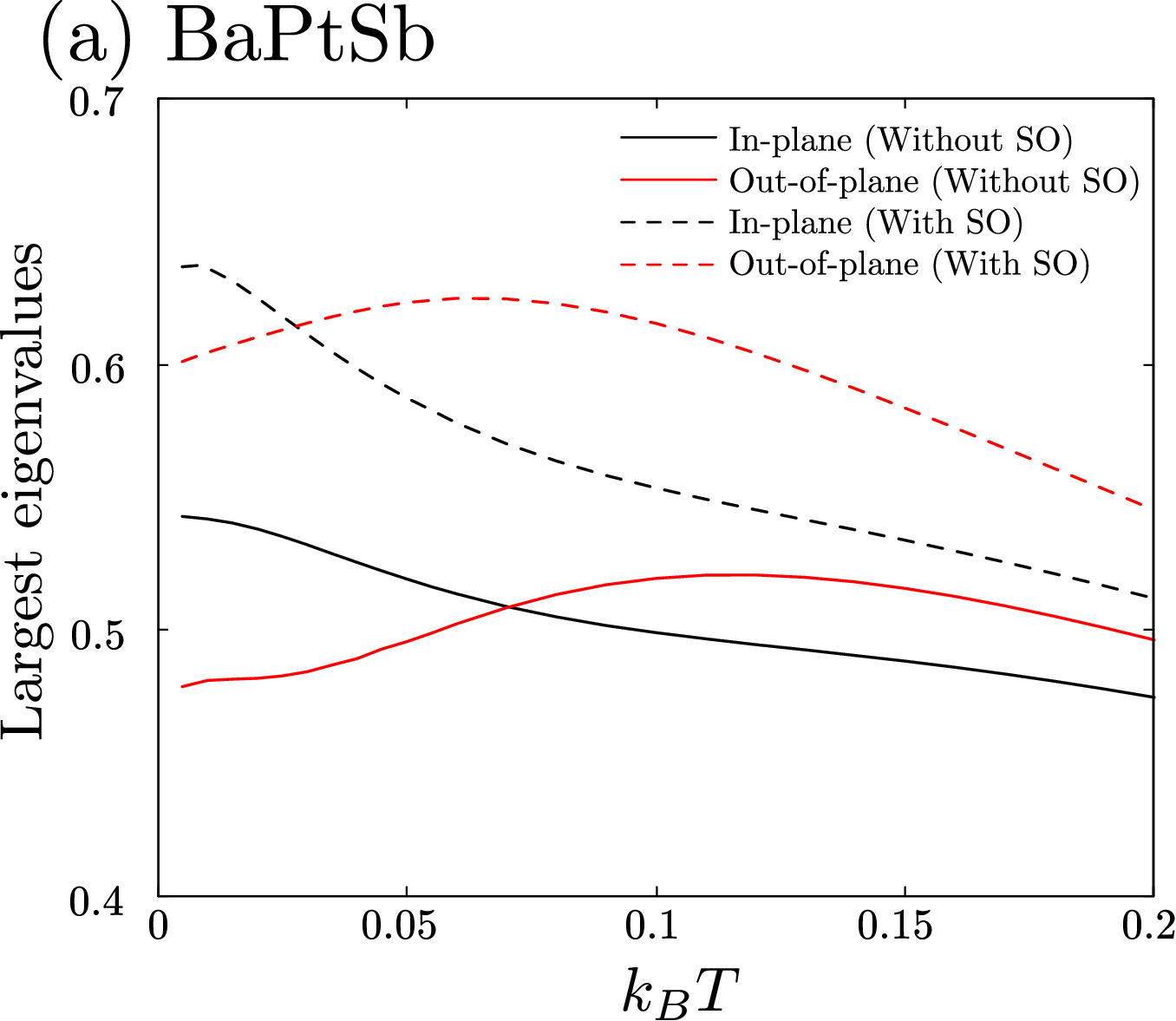}
	\includegraphics[width=40mm]{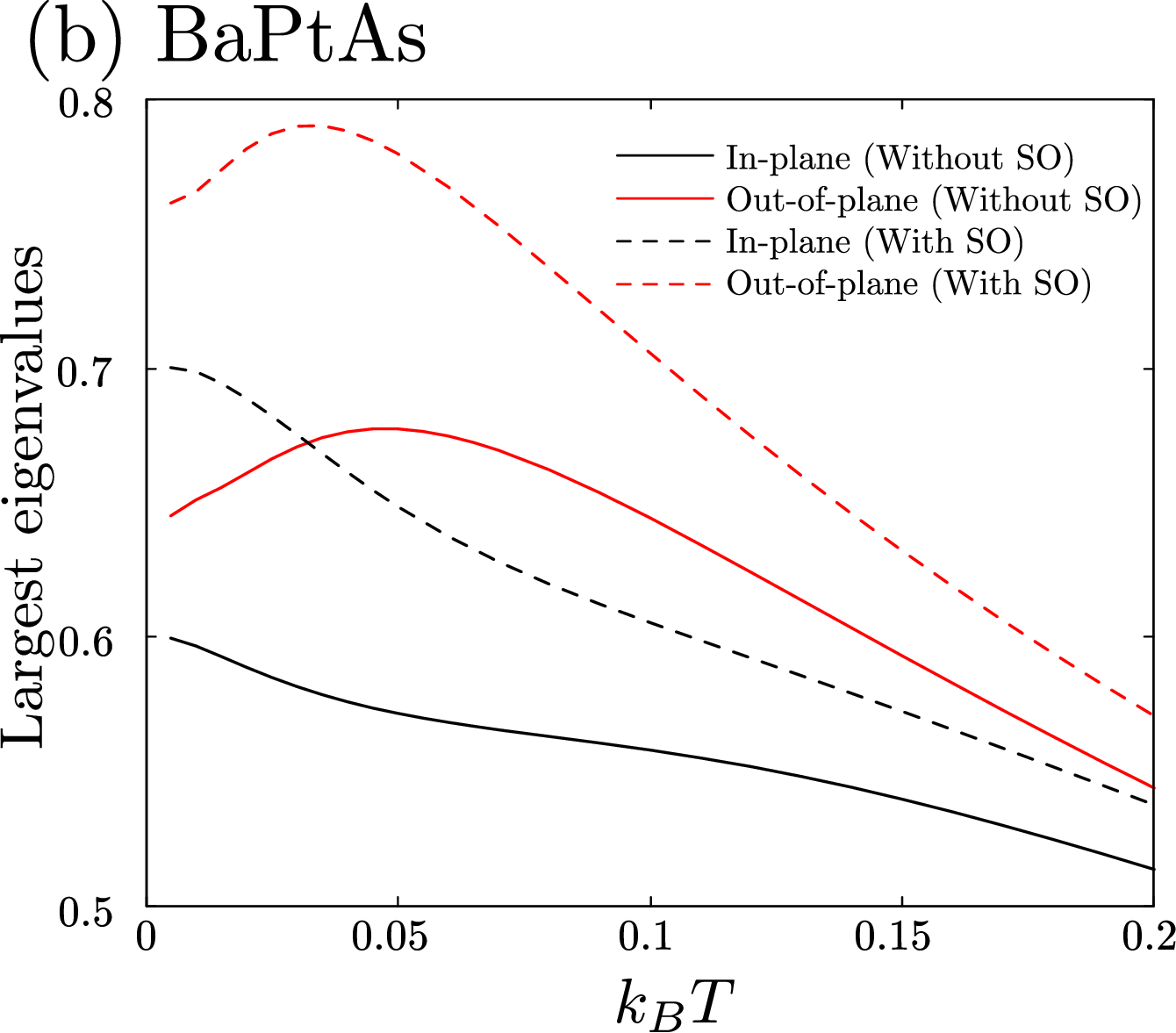}
	\caption{(Color online) Largest eigenvalues of static susceptibilities at the fixed $\bm q$ as a function of temperature; (a) BaPtSb and (b) BaPtAs. The black (red) lines denote the in-plane (out-of-plane) direction, and the solid (dashed) lines are results without (with) the SO interaction. }
	\label{chi_Tdep}
\end{figure}

Here, we study the temperature dependence of static susceptibilities. Figure \ref{chi_Tdep} displays the largest eigenvalues of static susceptibilities in the in-plane and out-of-plane directions at the fixed $\bm q$ as a function of temperature. 
The values of $\bm q$ in the in-plane direction in BaPtSb are given as $\approx (0, 0.7\pi,0)$ ($(0.06\pi, 0.9\pi,0)$) without (with) the SO interaction, and those in BaPtAs are $\approx (0.15\pi,0.6\pi,0)$ ($0.03\pi,0.7\pi,0)$ without (with) the SO interaction. The values of $\bm q$ in the out-of-plane direction are $(0,0,\pi)$ in all cases.
The peak positions in the in-plane direction slightly depend on the temperature.  The above-mentioned values of $\bm q$ in the in-plane direction give the maximum susceptibilities at $k_BT=0.005$ eV. 

The energy level of BaPtAs around the vHS is $\sim -0.1$ eV, which is farther from the Fermi level than that ($\sim -0.03$ eV) of BaPtSb. Except this low-energy region, the overall electronic structures in both BaPtSb and BaPtAs almost coincide.
Thus, there is no marked difference in susceptibilities between BaPtSb and BaPtAs in the higher-temperature region ($k_BT>\sim 0.1$ eV), where dominant fluctuations occur in the out-of-plane direction. Note that the peak position in the out-of-plane direction appears at $\bm q\simeq (0,0,\pi)$ in the whole temperature range.

On the other hand, since a better nesting condition exists in the in-plane direction around the vHS in BaPtSb, the maximum value of the static susceptibility in the in-plane direction is enhanced with decreasing temperature, and becomes dominant in the very low-temperature region. 

Since the SO interaction generates the splitting of Fermi surfaces, the nesting condition is weakened particularly in the in-plane direction. 
In-plane fluctuations in BaPtSb, however, become dominant at further lower-temperature region than the case without the SO interaction. 
While the amplitudes of static susceptibilities are rather sensitive to parameters in  effective models, the tendency of the temperature dependence of static susceptibility does not change. 

Although both compounds are very similar in the overall electronic structures and shapes of the Fermi surfaces, their dominant fluctuations differ. These results indicate that it is sensitive to the instabilities to ordered states in the in-plane (out-of-plane) direction, which might be associated with the occurrence of superconductivities in BaPtSb and BaPtAs.

\section{Summary}

We have investigated static susceptibilities in the new superconductors BaPtSb and BaPtAs with the honeycomb structure in the normal state.  Although the electronic structures obtained by the first-principles method in both compounds are generally very similar, the low-energy eigenvalues of BaPtSb around the M point with the van Hove singularity are much closer to the Fermi level than those of BaPtAs. We have also constructed the effective tight-binding models for both compounds with Pt 5$d$ $x^2-y^2$, $xy$ orbitals, and Sb 5$p$/As 4$p$ $z$ orbitals by using the Slater-Koster method, which capture features of the low-energy electronic structures and Fermi surfaces. 

We have calculated static susceptibilities on the basis of the effective models and found that the eigenstates corresponding to the largest eigenvalues of static susceptibilities originate from the outer spherelike Fermi surface around the K and K$^{'}$ points.  
Since the outer spherelike Fermi surface of BaPtSb in the $k_z=0$ plane has an equilateral triangular-like form, a better nesting condition exists in comparison with the BaPtAs case. 
In the low-temperature region, the maximum values of static susceptibilities appear in the in-plane direction in BaPtSb and in the out-of-plane direction in BaPtAs. 
We demonstrate that variations in the direction of fluctuations originate from the difference in low-energy electronic structures between these two compounds. 
Thus, to detect fluctuations, experimental studies of magnetic responses are expected. 

Since dominant fluctuations are strongly associated with the instabilities to ordered states, such as superconductivity, superconducting order parameters reflect these modes. 
In particular, the dominant mode in BaPtSb has an advantage in the occurrence of the TRSB superconducting state such as the $d_{x^2-y^2}+id_{xy}$ state. On the other hand, the dominant mode with $\simeq (0,0,\pi)$ might lead to the pairing wave functions with the horizontal node, which is different from the BaPtSb case. 
We will address in detail the properties including the pairing symmetry and transition temperature in the future.

Since the low-energy electronic structure of BaPtSb around the M point is very close to the Fermi level, the presence of external fields and/or perturbations can lead to the change in topology around the M point in the outer spherelike Fermi surface. 
This Lifshitz transition may affect topological aspects in the superconducting states with TRSB, and this is an interesting open problem.

\section*{Acknowledgment}
We are grateful to K. Kudo for many helpful discussions. This work was partially supported by JSPS KAKENHI Grant Numbers 19K03724 and 20K03826.

\bibliographystyle{jpsj}
\bibliography{paper}

\end{document}